\documentclass[aps,prx,twocolumn,superscriptaddress,floatfix,citeautoscript,preprintnumbers]{revtex4-2}
\usepackage{amsfonts, amssymb, amsmath, amsthm}
\usepackage{array}
\usepackage{bm}
\usepackage{braket}
\usepackage{dcolumn}
\usepackage{graphicx}
\usepackage{hyperref}
\hypersetup{
  colorlinks,%
  citecolor=blue,%
  filecolor=blue,%
  linkcolor=blue,%
  urlcolor=blue
}
\usepackage[normalem]{ulem}
\usepackage{dsfont}
\usepackage{indentfirst}
\usepackage{listings}
\usepackage{newtxtext,newtxmath}
\usepackage{subfigure}
\usepackage{float}
\usepackage{titlesec}
\usepackage{color}
\usepackage{xcolor}
\usepackage{enumitem}
\usepackage[title]{appendix}

\newtheorem{definition}{Definition}
\newtheorem{problem}{Problem}

\newcommand{\dInt}{{\mathrm{d}}}
\newcommand{\id}{{\mathds{1}}}
\newcommand{\tr}{{\mathrm{Tr}~}}

\newcommand{\varO}{\mathcal{O}}  

\usepackage{xparse,xcoffins}
\ExplSyntaxOn
\NewCoffin\imagecoffin
\NewCoffin\labelcoffin
\keys_define:nn { figs/label }
 {
  label   .tl_set:N = \l_label_tl,
  labelbox .bool_set:N = \l_label_box_bool,
  labelbox .default:n = true,
  fontsize .tl_set:N = \l_label_size_tl,
  fontsize .initial:n = \footnotesize,
  pos .choice:,
  pos/nw .code:n = \tl_set:Nn \l_label_pos_tl { left,up },
  pos/ne .code:n = \tl_set:Nn \l_label_pos_tl { right,up },
  pos/sw .code:n = \tl_set:Nn \l_label_pos_tl { left,down },
  pos/se .code:n = \tl_set:Nn \l_label_pos_tl { right,down },
  pos/n .code:n = \tl_set:Nn \l_label_pos_tl { hc,up },
  pos/w .code:n = \tl_set:Nn \l_label_pos_tl { left,vc },
  pos/s .code:n = \tl_set:Nn \l_label_pos_tl { hc,down },
  pos/e .code:n = \tl_set:Nn \l_label_pos_tl { right,vc },
  pos .initial:n = nw,
  unknown .code:n   = \clist_put_right:Nx \l_label_clist
                       { \l_keys_key_tl = \exp_not:n { #1 } }
 }
\clist_new:N \l_label_clist
\box_new:N \l_label_box
\box_new:N \l_label_image_box

\NewDocumentCommand{\xincludegraphics}{O{}m}
 {
  \group_begin:
  \tl_clear:N \l_label_tl
  \clist_clear:N \l_label_clist
  \keys_set:nn { figs/label } { #1 }
  \tl_if_empty:NTF \l_label_tl
   {
    \includegraphics:Vn \l_label_clist { #2 }
   }
   {
    \SetHorizontalCoffin\imagecoffin
     {
      \includegraphics:Vn \l_label_clist { #2 }
     }
    \SetHorizontalCoffin\labelcoffin
     {
      \raisebox{\depth}
       {
        \bool_if:NTF \l_label_box_bool
         { \fcolorbox{white}{white}{\l_label_size_tl\l_label_tl} }
         { \l_label_size_tl\l_label_tl }
       }
     }
    \SetVerticalPole\imagecoffin{left}{3pt+\CoffinWidth\labelcoffin/2}
    \SetVerticalPole\imagecoffin{right}{\Width-3pt-\CoffinWidth\labelcoffin/2}
    \SetHorizontalPole\imagecoffin{up}{\Height-3pt-\CoffinHeight\labelcoffin/2}
    \SetHorizontalPole\imagecoffin{down}{3pt+\CoffinHeight\labelcoffin/2}
    \use:x{\JoinCoffins\imagecoffin[\l_label_pos_tl]\labelcoffin[vc,hc]} 
    \TypesetCoffin\imagecoffin
   }
   \group_end:
 }
\NewDocumentCommand{\setlabel}{m}
 {
  \keys_set:nn { figs/label } { #1 }
 }
\cs_new_protected:Nn \includegraphics:nn
 {
  \includegraphics[#1]{#2}
 }
\cs_generate_variant:Nn \includegraphics:nn { V }
\ExplSyntaxOff

\begin{document}

\title{Simulating prethermalization using near-term quantum computers}

\begin{abstract}
    Quantum simulation is one of the most promising scientific applications of quantum computers. Due to decoherence and noise in current devices, it is however challenging to perform digital quantum simulation in a regime that is intractable with classical computers. In this work, we propose an experimental protocol for probing dynamics and equilibrium properties on near-term digital quantum computers. As a key ingredient of our work, we show that it is possible to study thermalization even with a relatively coarse Trotter decomposition of the Hamiltonian evolution of interest. Even though the step size is too large to permit a rigorous bound on the Trotter error, we observe that the system prethermalizes in accordance with previous results for Floquet systems. The dynamics closely resemble the thermalization of the model underlying the Trotterization up to long times. We extend the reach of our approach by developing an error mitigation scheme based on measurement and rescaling of survival probabilities. To demonstrate the effectiveness of the entire protocol, we apply it to the two-dimensional XY model and numerically verify its performance with realistic noise parameters for superconducting quantum devices. Our proposal thus provides a route to achieving quantum advantage for relevant problems in condensed matter physics.
\end{abstract}

\author{Yilun Yang}
\author{Arthur Christianen}
\author{Sandra Coll-Vinent}
\affiliation{Max-Planck-Institut f\"{u}r Quantenoptik, 85748 Garching, Germany}
\affiliation{Munich Center for Quantum Science and Technology, 80799 M\"{u}nchen, Germany}

\author{Vadim Smelyanskiy}
\affiliation{Google Quantum AI, Venice, California 90291, USA}

\author{Mari Carmen Bañuls}
\affiliation{Max-Planck-Institut f\"{u}r Quantenoptik, 85748 Garching, Germany}
\affiliation{Munich Center for Quantum Science and Technology, 80799 M\"{u}nchen, Germany}

\author{Thomas E.~O'Brien}
\affiliation{Google Quantum AI, 80636 M\"{u}nchen, Germany}

\author{Dominik S.~Wild}
\author{J.~Ignacio Cirac}
\affiliation{Max-Planck-Institut f\"{u}r Quantenoptik, 85748 Garching, Germany}
\affiliation{Munich Center for Quantum Science and Technology, 80799 M\"{u}nchen, Germany}

\date{\today}							
\maketitle

\newpage
\section{Introduction}

Quantum computers promise to have a great impact on scientific research. A particular example is the study of thermalization of quantum many-body systems. The problem is computationally challenging with classical methods~\cite{Vidal2004, Ido2015, Carleo2017} as it requires simulating the long-time dynamics of large systems. A fault-tolerant quantum computer would render this problem tractable by enabling quantum simulation~\cite{Feynman1982, Lloyd1996, Georgescu2014, Altman2021}. 

Despite impressive recent progress, present day quantum devices are still far from the regime of fault tolerance. Any current quantum simulation is therefore affected by noise and imperfections. In circuit-based quantum computers, continuous-time dynamics can be approximated using, for example, Trotterization~\cite{childs2021}. With state-of-the-art gate errors~\cite{Yoneda2018a, Arute2019, Huang2019a, Wu2021, Mi2022b, Wei2022a}, it is however only possible to run simulations with a controlled Trotter error up to short times, which are insufficient to explore thermalization in classically intractable systems (50 or so qubits in two or more dimensions). 

\begin{figure}[b]
    \centering
    \includegraphics[width=0.45\textwidth]{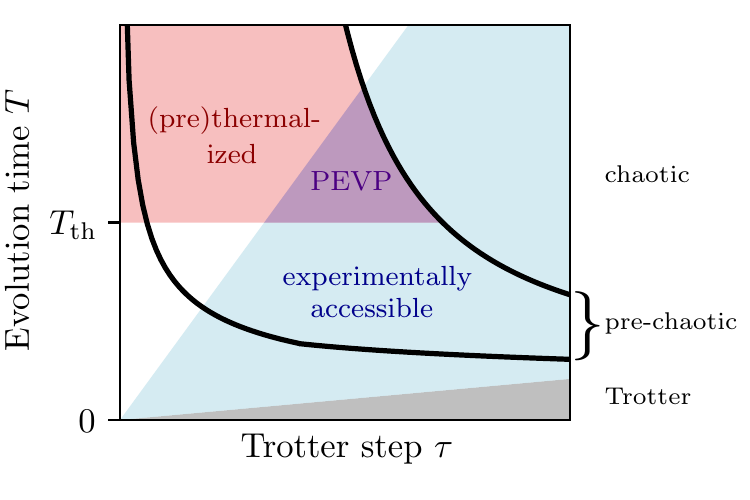}
    \caption{
        Different regimes of the dynamics of local observables depending on the Trotter step $\tau$ and the evolution time $T$. The black lines separate three regimes: bounded Trotter error (bottom), Floquet prethermalization (middle), and chaotic dynamics (top). The lower line scales as $\mathcal{O}\left( \max\left\{ \tau^{- p/(d+1)}, N^{-1}\tau^{-p} \right\} \right)$, following error bounds for the $p^\mathrm{th}$-order Trotter decomposition with system size $N$. The upper line is determined by the Floquet heating time and scales as $e^{\mathcal{O}\left( 1/\tau \right)}$. The blue shaded area indicates constant maximum circuit depth, relevant for noisy quantum computers. The grey area is excluded due to the constraint that $T \ge \tau$. The red shading highlights where the total time $T$ exceeds a system-dependent (pre)thermalization time scale $T_{\mathrm{th}}$. The prethermalized expectation value problem is experimentally accessible in the purple intersection.
    }
\label{fig:scheme}
\end{figure}

In this work, we demonstrate that thermalization can already be observed for much larger Trotter steps than needed to guarantee a bounded Trotter error, making it feasible to study this phenomenon on near-term quantum devices. In this regime, the system may be viewed as subject to a periodic Floquet drive \cite{Goldman2014,Bukov2015,Moessner2017}, where one Trotter step corresponds to one period. The fate of Floquet systems at late times has been a subject of recent interest~\cite{Sieberer2019, Kargi2021,Morningstar2022}. Even though the system generally heats up to infinite temperatures ~\cite{lazarides2014,dalessio2014}, the heating time may be very long if the driving frequency is large compared to all local energy scales \cite{abanin2015}. The system then \emph{prethermalizes} ~\cite{kuwahara2016,Mori2016, Else2017, Mori2018, Pizzi2021, Ye2021}: Before it heats up, its dynamics mirror the equilibration of a closed system. 
The prethermal regime is relatively easy to access in practice because the Floquet heating time increases exponentially with the driving frequency or, equivalently, the inverse Trotter step size (see Fig.~\ref{fig:scheme}).

With this in mind, we define the \emph{prethermalized expectation value problem} (PEVP): Given a Floquet unitary and a product initial state,  what value does a local observable reach in the prethermal plateau? We find that this problem can be solved even in presence of realistic noise. Following a small circuit adjustment, the PEVP turns out to be amenable to a simple but highly effective error-mitigation scheme based on rescaling survival probabilities. Using this strategy, the error-mitigated PEVP reproduces the equilibrium properties of a model that is closely related to the Hamiltonian underlying the Trotterization. More precisely, the prethermal expectation values describe the diagonal ensemble of this model, which is equivalent to the microcanonical ensemble assuming that the eigenstate thermalization hypothesis (ETH)~\cite{Srednicki1999, Rigol2008a, DAlessio2016, Deutsch2018} is valid. Besides its application to the study of thermalization, the PEVP may be viewed as a problem of independent computational interest in the context of demonstrating quantum advantage.

The paper is structured as follows. In Sec.~\ref{sec:pre}, we discuss thermalization in Floquet systems and present simulation results for the two-dimensional XY model as an example. We introduce our error mitigation strategy based on the rescaling of survival probabilities in Sec.~\ref{sec:mitigation}, where we also provide a thorough numerical analysis of its performance. Equipped with that, we demonstrate the suitability of the PEVP for near-term devices by simulating it with realistic noise parameters of superconducting quantum computers. We conclude in Sec.~\ref{sec:summary}.
\section{The prethermalized expectation value problem}
\label{sec:pre}

\subsection{Time evolution on digital quantum computers\label{sec:equilibration}}
The time evolution under a Hamiltonian $H$ can be reproduced on a digital quantum computer using the Suzuki--Trotter decomposition. In its simplest, first-order form, the decomposition approximates the time-evolution unitary $U(\tau) = e^{- i H \tau}$ by
\begin{eqnarray}\label{def:Utrot}
	U_{\mathrm{Trotter}}(\tau) = \prod_{j=1}^{\Gamma} e^{-i H_j\tau}.
\end{eqnarray}
where $H = \sum_{j=1}^{\Gamma} H_j$. Each $H_j$ is a sum of mutually commuting local terms, such that $e^{-i H_j \tau}$ can be efficiently implemented using local gates. The smaller the Trotter step $\tau$, the more accurate the Trotter decomposition. For the $p$-th order Trotter decomposition \cite{Hatano2005}, which generalizes the previous simple formula, the error of $U_\mathrm{Trotter}(\tau)$ with respect to the desired unitary $U(\tau)$ is bounded from above by $\varO(N\tau^{p+1})$, where $N$ is the system size~\cite{childs2021}. The dependence on $N$ can be eliminated if all quantities of interest are local observables. According to the Lieb--Robinson bound, only a light cone with a radius proportional to the total evolution time $T$ is relevant~\cite{Lieb1972}. Therefore, the system size $N$ can be replaced with the size of the light cone $\sim T^d$ before it reaches the edges of the system, where $d$ is the spatial dimension. We hence require that the Trotter step $\tau$ be less than $\varO(\max\left\{T^{-(d+1)/p}, (NT)^{-1/p}\right\})$ for the Trotterized time evolution of local observables to converge to the continuous evolution under $H$.

We can now define the following computational problem.
\begin{problem}[The Trotter time-average problem]
    \label{def:time_average}
    Given a unitary $U_{\mathrm{Trotter}}(\tau)$, a state $\ket{\psi}$, a local observable $A$ and a time $t=m \tau$ for positive integer $m$, and a small positive constant $\epsilon$, 
    compute the time-averaged observable
\begin{eqnarray} \label{eq:AFloq}
    \braket{A}_t = \frac{1}{m + 1} \sum_{n = 0}^{m}  \braket{ \psi |  U^{\dagger}_{\mathrm{Trotter}}(\tau)^{n} A
    U_{\mathrm{Trotter}}(\tau)^{n} | \psi}
\end{eqnarray}
within additive error $\epsilon \Vert A\Vert$, where $\Vert \cdot \Vert$ is the operator norm.
\end{problem}
Note that the Trotterization is not uniquely defined by the Hamiltonian and $U_\mathrm{Trotter}$ must be specified explicitly. The cost of solving this problem on a classical computer generically scales exponentially with either the number of Trotter steps $m$ or the system size $N$~\footnote{For example, a state vector simulation scales linearly in the number of Trotter steps but exponentially with the system size. While a tensor network simulation scales polynomially in system size but exponentially with the number of Trotter steps.}, whereas on a fault-tolerant quantum computer, the effort increases at most polynomially with both. The hardness of the problem is further supported by the fact that it becomes BQP-complete at times $t = \mathrm{poly}(n)$ if the Trotter error is negligible~\cite{janzing2005}. In section~\ref{sec:mitigation}, we present evidence that the problem is solvable on noisy quantum computers up to a maximum number of Trotter steps, which is independent of system size. We then show in section~\ref{sec:implementation} that noisy quantum devices may reach a classically intractable regime with realistic noise parameters, even when taken into account the overhead of our error mitigation strategy.

\subsection{Prethermalization}

Problem~\ref{def:time_average} is not only interesting from the perspective of dynamics but it can also yield insight into equilibrium properties. In condensed matter or statistical physics, one would typically describe a system in equilibrium in terms of its temperature, or in case of the microcanonical ensemble, its internal energy. Under ETH, 
the microcanonical ensemble at the mean energy of the state $\ket{\psi}$ can be approximated by solving Problem~\ref{def:time_average}.

More precisely, in the limit of continuous time evolution, the long-time average of an observable is described by the diagonal ensemble. For a given initial state $\ket{\psi}$ and an observable $A$,
\begin{eqnarray}
    \begin{aligned}
        \lim_{T\to\infty}\frac{1}{T} \int_0^{T}  \braket{\psi (t)| A | \psi (t)} \dInt t
        = \sum_{k} |\braket{k | \psi}|^2 \braket{k | A | k},
    \end{aligned}
\end{eqnarray}
where $H = \sum_{k} E_k \ket{k}\bra{k}$ is the spectral decomposition of a non-degenerate Hamiltonian~\footnote{In the case of degenerate Hamiltonian spectrum, one can still diagonalize the observable projected onto each subspace of Hamiltonian eigenvalue to define the diagonal ensemble as long time average}. Assuming ETH, the expectation value $\braket{k | A | k}$ is a smooth function of the energy $E_k$ up to a small, state-dependent correction~\cite{Srednicki1999}. The diagonal ensemble is then equivalent to the microcanonical ensemble at energy $\braket{\psi | H | \psi}$ provided the energy variance of $\ket{\psi}$ is sufficiently small. For observables that are an average of an extensive number of local terms, e.g., the total magnetization per site, we expect the microcanonical ensemble to vary significantly only on an extensive energy scale. It is thus possible to estimate expectation values in the microcanonical ensemble from the diagonal ensemble of states whose width in energy is subextensive. Product states satisfy this condition as their widths in energy are (under weak assumptions) proportional to $\sqrt{N}$~\cite{Hartmann2004}.

The above discussion shows that it is possible to probe the microcanonical ensemble by solving problem~\ref{def:time_average} with product initial states at different mean energies. This is, however, challenging with current quantum devices for two reasons. First, the maximum number of Trotter steps $T/\tau$ is limited by the maximum circuit depth in the presence of noise, while the total time $T$ required to reach equilibrium may be large. Therefore, noisy quantum devices are usually unable to reach long enough times with bounded Trotter error.  Secondly, the finite calibration precision renders it challenging to get high relative precision in the angle of rotation for gates that are very close to the identity, bounding from below the size of $\tau$. 

We will now argue that it is nevertheless possible to study equilibrium phenomena. Using larger, experimentally feasible Trotter steps can be viewed as applying a periodic Floquet drive. The system can be described by the Floquet Hamiltonian $H_F$, which is implicitly defined by
\begin{eqnarray}
	U_{\mathrm{Trotter}}(\tau) = e^{-i H_F \tau}.
\end{eqnarray}
The Floquet Hamiltonian is not unique as its eigenvalues are only defined modulo $\omega=2 \pi / \tau$, the effective driving frequency. For large $\tau$, (small $\omega$), i.e., outside the Trotter limit, the Floquet Hamiltonian is highly non-local and will cause a generic initial state to heat up to infinite temperature~\cite{lazarides2014,dalessio2014}. Despite this, it is possible to observe (approximate) equilibration if the heating time scale is much greater than the equilibration time scale. This is known as Floquet prethermalization~\cite{kuwahara2016, Fleckenstein2020, Morningstar2021}. Fortunately for our purposes, Floquet prethermalization is relatively easy to access because Floquet heating occurs on a time scale $t_F \propto e^{\varO(\omega / kJ)}$, where $k$ is the interaction range and $J$ is the local energy scale, assuming $\omega \gtrsim kJ $. We highlight the favorable exponential dependence of $t_F$ on $\omega / k J$ and the fact that $k J$ is independent of the system size.

For times much less than $t_F$, the system evolves approximately according to an effective Hamiltonian which is close to, but not the same as, the original Hamiltonian $H$. More precisely, the effective Hamiltonian is local and it is given by the $n_0$-th order Magnus expansion~\cite{Magnus1954, Blanes2009} of the Floquet Hamiltonian, where $n_0 = \varO(\omega / kJ)$ (see Appendix~\ref{sec:magnus} for details). Observables start to equilibrate under the effective Hamiltonian before eventually heating up. If the equilibration time $t_0$ is much shorter than $t_F$, then there exists a prethermal plateau $t_0 \le t \ll t_F$, during which the expectation value of the observable is approximately constant. We provide a formal definition of a plateau in Appendix \ref{sec:def}. 

The above observations motivate the definition of the PEVP: 
\begin{problem}[Prethermalized expectation value problem]
    \label{prob:PEVP}
    Given a unitary $U_{\mathrm{Trotter}}(\tau)$, a state $\ket{\psi}$, and a local observable $A$, assume that a prethermal plateau exists between times $t_1$ to $t_2$, such that $\max_{t \in [t_1, t_2)} \langle A \rangle_t - \min_{t \in [t_1, t_2)} \langle A \rangle_t \leq \epsilon \Vert A\Vert$ for some positive constant $\epsilon$. Find the value of $\braket{A}_t$ to within additive error $2 \epsilon \Vert A\Vert$ for any $t \in [t_1, t_2)$ .
\end{problem}
This problem reduces to solving Problem~\ref{def:time_average} at time $t = t_1$. In the following sections, we show using the example of the two-dimensional XY model that the prethermal plateau is indeed accessible and that the properties of the effective Hamiltonian closely resemble those of the initial Hamiltonian. We further demonstrate that the PEVP can be solved on a noisy quantum device with realistic parameters up to system sizes for which classical simulation of the dynamics is intractable.

\subsection{PEVP with the XY model\label{sec:xy}}
We focus on the two-dimensional quantum XY model on a square lattice for the remainder of this work. We emphasize, however, that the approach can be readily applied to many other models. The Hamiltonian of the XY model is given by
\begin{eqnarray}
	H_{\mathrm{XY}} = - J \sum_{\braket{ij}} \left( S^x_i S^x_j + S^y_i S^y_j \right),
\end{eqnarray} 
where $J$ is the interaction strenth, $S_i^\alpha$ ($\alpha \in \{ x, y, z\}$) are spin-1/2 operators on site $i$, and the sum runs over all pairs of nearest neighbors. The model is convenient for digital quantum computers as its two-site interaction generates a partial iSWAP gate,
\begin{eqnarray}
    e^{-i J\left( S^x_i S^x_j + S^y_i S^y_j \right) \tau} = \text{iSWAP}^{ - J \tau / \pi}_{ij}.
\end{eqnarray}
A single Trotter step in a first-order decomposition consists of applying a partial iSWAP gate to each nearest-neighbor pair of qubits. As non-overlapping gates can be performed in parallel, these operations can be carried out in a circuit whose depth is equal to the number of nearest neighbors (4 in the case of the square lattice).

The XY model in two dimensions can be solved with quantum Monte Carlo algorithms~\cite{Loh1985, Ding1992} and thus serves as a good benchmark to our method. It is known to undergo the Kosterlitz--Thouless (KT) transition~\cite{Kosterlitz1973, Ding1992} at nonzero temperature. This phase transition can be characterized by the mean-squared in-plane magnetization per site,
\begin{eqnarray}
	m_{x}^2 + m_{y}^2 = 4\cdot \frac{ \left( \sum_i S^x_i\right)^2 + \left( \sum_i S^y_i\right)^2 }{N^2},
\end{eqnarray}
which is an approximation to the in-plane susceptibility~\cite{Ding1992}. The mean-squared magnetization can be written as the sum of two-site correlators, which decay exponentially with the distance between the two sites at high temperature. Hence, $m_x^2 + m_y^2$ decreases with the system size as $1/N$ in the thermodynamic limit. Below the critical temperature, the system exhibits quasi long-range order. The mean-squared magnetization decays only as $1/N^{1/8}$ and its value remains non-negligible for moderately large systems ~\cite{Ding1992}.

\begin{figure}
    \centering
    \xincludegraphics[width=0.45\textwidth, label=(a)]{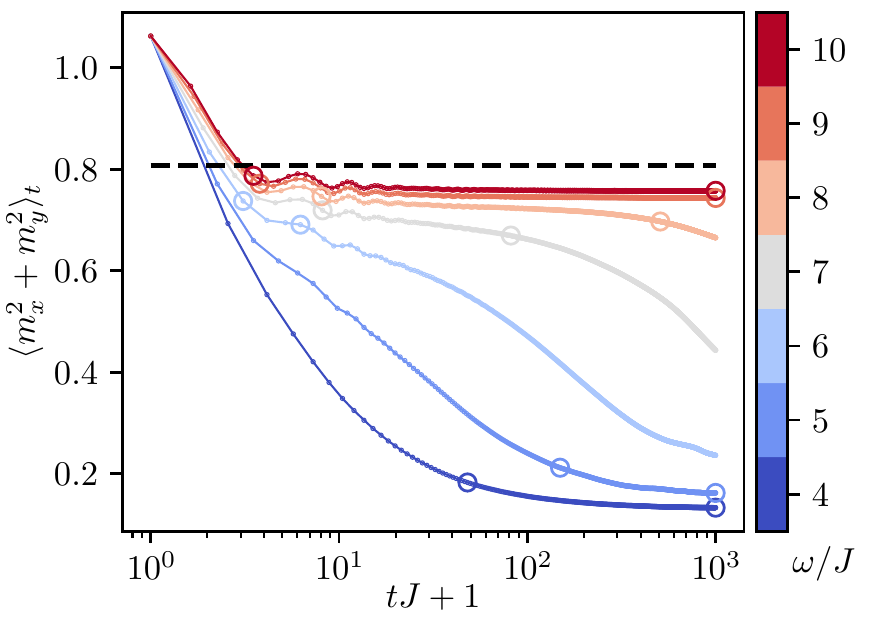}
    \xincludegraphics[width=0.48\textwidth, label=(b)]    {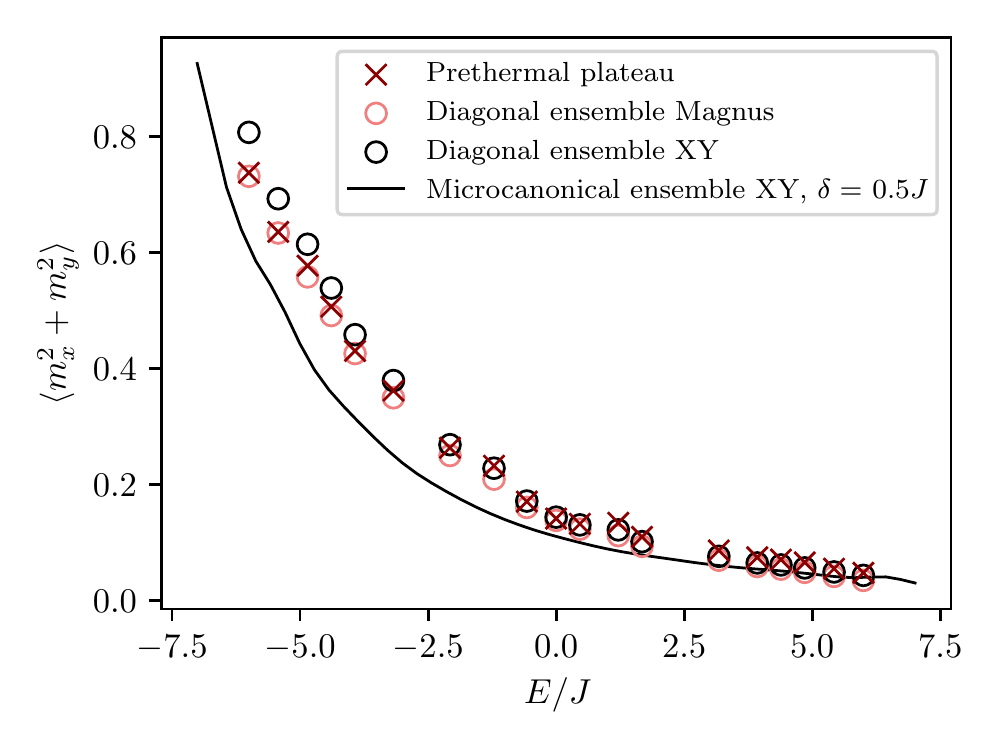}
    \caption{\textbf{(a)}~Prethermal plateau of the 2D XY model for system size $N = 4 \times 4$. The initial state is $\ket{X+}$. The colored lines show the time averages of the mean-squared in-plane magnetization for different Trotter step sizes $\tau$, corresponding to different driving frequencies $\omega = 2 \pi / \tau$. The large circles stand for the starting and end points of the plateaus according to Definition~\ref{def:plateau} with tolerance $\epsilon = 0.05$ and a maximum value of $t_2 J$ of $10^3$. The black dashed line represents the value in the diagonal ensemble of the initial Hamiltonian. \textbf{(b)}~Comparison of the value at the prethermal plateau with the values in the microcanonical and diagonal ensemble values of the initial XY Hamiltonian and in the diagonal ensemble of the first-order Magnus expansion. The system size is $4 \times 4$. The driving frequency is as $\omega = 8J$ and the plateau value is taken from the time average at $t = 20 / J$, which is on the prethermal plateau for all computed initial states with tolerance $\epsilon = 0.05$. For the microcanonical ensemble, we average over an energy window of width $\delta = 0.5J$ in the $m_z = 0$ subspace (see Appendix~\ref{sec:def}).}
    \label{fig:prethermal}
\end{figure}

In analogy to the long-time average that gives rise to the diagonal ensemble, we probe the prethermal plateaus using the Floquet time average as in Definition~\ref{def:time_average}, where the Trotterization is shown in the appendix in Fig.~\ref{fig:magnus}a. We explore this quantity using exact diagonalization on a square lattice with $N = 4 \times 4$ spins and open boundary conditions. Figure~\ref{fig:prethermal}a shows the values of the mean-squared in-plane magnetization for the initial state $\ket{\psi} = \ket{X+} = \left[ \frac{1}{\sqrt{2}}\left( \ket{0} + \ket{1} \right)\right]^{\otimes N}$. The different colors indicate the Trotter step size $\tau$ or, equivalently, the driving frequency $\omega = 2\pi / \tau$. The initial state is close to the ground state of the XY Hamiltonian. We therefore expect the in-plane magnetization to remain high in the prethermal plateau, provided the effective Hamiltonian does not differ too much from the XY model.

We indeed observe prethermal plateaus for large driving frequencies ($\omega \ge 8J$), and these last for $t > 10^3 / J$ when $\omega \ge 9J$. The plateau values approach the diagonal ensemble value (black dashed line) with increasing driving frequencies. They deviate only slightly due to the correction in the Magnus expansion, which will be discussed later in this subsection. This confirms that the dynamics with fast Floquet drive are similar to the dynamics of the original Hamiltonian in this prethermal regime. By contrast, no plateaus are observed at low driving frequencies, where the time average of the mean-squared magnetization quickly drops to expected value at infinite temperature, $2 / N$. 

We may perform the same analysis for different initial states. We choose product states in which the spins on the two sublattices of the square lattice are in the respective states $\ket{\theta, 0}$ and $\ket{\pi - \theta, \phi}$, where $\ket{\theta, \phi} = \cos(\theta / 2) \ket{0} + \sin(\theta /2 ) e^{i\phi} \ket{1}$ parametrizes an arbitrary state of a qubit (spin-1/2). This choice of states allows us to cover a wide range of the spectrum while ensuring that the total magnetization in the $z$ direction vanishes. The latter constraint is convenient because the Hamiltonian conserves the total $z$-magnetization, $m_z = \sum_{i = 1} ^{N} \sigma^z_i / N$. Thermalization therefore occurs in the eigenspaces of $m_z$. Low-energy product states however are not eigenstates of $m_z$. By choosing the expectation value of $m_z$ to be zero, we maximize the overlap of the product state with the sectors of low $z$-magnetization, for which we expect similar equilibration dynamics.

We find that all product states of the above form exhibit prethermal plateaus at similar driving frequencies and evolution times. We evaluate the prethermal values of the in-plane magnetization by performing the Floquet time average up to time $t = 20/J$ with driving frequency $\omega = 8 J$. The result is shown for various initial states as a function of their mean energy in Fig.~\ref{fig:prethermal}b. For comparison, we also show the diagonal and microcanonical ensemble values of the initial XY model, as well as the diagonal ensemble one of the first-order Magnus expansion of Floquet Hamiltonian, given by
\begin{eqnarray}
	\begin{aligned}
		H_{\mathrm{Magnus}}^{(1)} =  & \frac{1}{\tau} \int_{0}^{\tau} \dInt t_1 H(t_1)            \\
		& + \frac{1}{2i \tau} \int_{0}^{\tau} \dInt t_1 \int_{0}^{t_1} \dInt t_2 \left[H(t_1), H(t_2)\right].
	\end{aligned}
\end{eqnarray}
Here, $H(t)$ is the piecewise constant Hamiltonian corresponding to the different terms of the Trotter expansion Eq.~(\ref{def:Utrot}): 
\begin{eqnarray}
    H(t) = \Gamma H_j \text{ for } (j-1)\tau / \Gamma \le t < j \tau / \Gamma,
\end{eqnarray}
where $1 \le j \le \Gamma$. Definitions of the different ensembles and higher orders of the Magnus expansion can be found in App.~\ref{sec:def} and App.~\ref{sec:magnus}, respectively.

The values at the prethermal plateau are close to those of the diagonal ensemble $H_\mathrm{Magnus}^{(1)}$, indicating that the first-order truncation already serves as a good approximation for Floquet Hamiltonian in the prethermal regime. In Appendix~\ref{sec:magnus}, we show that the higher orders lead to no significant improvement for $\omega = 8J$. The thermal equilibrium values of the initial XY Hamiltonian, in both the diagonal and the microcanonical ensemble, deviate slightly from the Floquet values. Nevertheless, the comparison indicates that the prethermal properties of the Floquet system can reveal nontrivial thermal properties of the XY Hamitlonian.

\section{Error mitigation}
\label{sec:mitigation}

\subsection{Rescaling of survival probabilities}

Without mitigation, noise will frustrate any naive attempts to observe prethermal plateaus on current quantum hardware. As we show in Appendix~\ref{sec:diff_TE}, noise provides an additional heating source to the Floquet driving already discussed; one that we expect to be far stronger with today's error rates, and one without favourable scaling in the system size. It is therefore desirable to develop an error mitigation technique to estimate the result of a noiseless quantum circuit from multiple measurements in a noisy circuit~\cite{Temme2017, Endo2018a, Cai2022}. However, we do not see a reliable method for extracting the desired noiseless results from measurements of the noisy state as this would imply the ability of inferring low-temperature results from high-temperature ones.

To circumvent this issue, we avoid direct tomography of the time-evolved observables on the noisy state. Instead, we convert observable estimation into a survival probability circuit, in a manner similar to that used in out-of-time-order correlators (OTOC)~\cite{mi2021} or echo verification circuits~\cite{obrien2021,huo2022}. Following forward evolution, we \emph{apply} the observable and then evolve backwards in time, followed by a projection onto the initial state (see Fig.~\ref{fig:scaling}a). This yields a survival probability of the form
\begin{eqnarray}
    L_{A, \psi}(t) = \left| \braket{\psi | e^{iHt} A e^{-i H t} | \psi} \right|^2 = \braket{\psi | A(t) | \psi}^2.
\end{eqnarray}
In the following, we drop the label $\psi$ for notational simplicity. For this procedure to work, $A$ must be a (local) unitary. For spin systems, it is possible to write any observable as a sum of products of unitary Pauli operators and to measure each Pauli operator separately. Although $L_{A}(t)$ only gives the expectation value of an observable up to a sign, one can infer the sign by tracking it from the known initial value, assuming $\braket{\psi | A(t) | \psi }$ is a smooth function~\cite{Lu2020}.
This simplifies previous Loschmidt-echo style methods for learning $\braket{\psi | A(t) |\psi}$, which required ancilla qubits, the preparation of large Greenberger-Horne-Zeilinger (GHZ) states~\cite{obrien2021} or intermediate re-preparation and measurement of qubits~\cite{huo2022}.

As we will now demonstrate, a simple rescaling is remarkably effective at mitigating errors in the estimation of the survival probability. The strategy is based on the observation that the survival probability is approximately proportional to the probability of no error occurring. The reason is that the state becomes highly entangled during the evolution, at which point a single-qubit error results in an orthogonal state with high probability. To be more concrete, consider a single Pauli error $\sigma^{\mu}_i$ occurring at time $t^{\prime} < t$ at site $i$ and set the observable $A$ to be identity. The survival probability is then given by $[\tr (\rho_i(t^{\prime}) \sigma^{\mu}_i)]^2$, where $\rho_i(t^{\prime})$ is the reduced density matrix of $\ket{\psi(t^{\prime})}$ at site $i$. If this site is entangled with the other parts of the system, the reduced density matrix will be close to the identity (completely mixed) and the survival probability will be close to zero. 

The above discussion suggests that the survival probability with noise is related to the noiseless value, times the probability that no error has occurred. For concreteness, we consider error models in which a single-qubit noise channel $\mathcal{N}_p$ is applied to each qubit after every layer of unitary gates. Here, $p$ is the probability that the channel causes an error on the qubit. The state of art gate error rate is around $0.5\%$ for two-qubit gates~\cite{Mi2022b, Wei2022a}, motivating our choice of $p = 0.3\%$ per qubit per gate as the reference value in our model~\footnote{In experiments, XY rotations are sometimes compiled into more than one two-qubit gate. The value of $p$ should then be increased accordingly.}.

Denoting the survival probability in the presence of noise by $L_A^{\mathcal{N}_p}(t)$, we then expect that
\begin{eqnarray}
    L_{A}^{\mathcal{N}_p}(t) / L_{A}(t) \approx (1 - p)^{ND},
    \label{eq:scaling}
\end{eqnarray}
where $N$ is the number of qubits and $D$ is the circuit depth including both forward and backward evolutions. Crucially, no independent knowledge of the noise channel is required to estimate $L_A(t)$. By setting $A = \id$, we obtain $L_\id^{\mathcal{N}_p}(t) \approx (1 - p)^{ND}$ since the noiseless survival probability satisfies $L_\id(t) = 1$. Hence,
\begin{eqnarray}
    L_A(t) \approx L_A^{\mathcal{N}_p}(t) / L_\id^{\mathcal{N}_p}(t),
    \label{eq:rescale}
\end{eqnarray}
where the right-hand side can be obtained from measurements on the noisy quantum device.

We can make this argument more rigorous for channels that can be represented in terms of unitary Kraus operators. For such channels, the probability that a particular error occurs is independent of the state. This class of channels includes depolarizing and dephasing noise as well as all other Pauli channels~\footnote{Even though amplitude damping error is not included in this class of channels, we find that the conclusions of this section nevertheless hold to a good approximation. See Appendix~\ref{sec:supp_plot} for numerical results.}. The survival probability after the noisy circuit can be expressed as
\begin{eqnarray}
    L_{A}^{\mathcal{N}_p}(t) = \tr \left[ \left( A \rho^{\mathcal{N}_p}_{\psi}(t) \right)^2\right],
    \label{eq:noisy_sp}
\end{eqnarray}
where $\rho^{\mathcal{N}_p}_{\psi}(t)$ is the mixed state after the noisy forward evolution~\footnote{To obtain this equation, the circuit in Fig.~\ref{fig:scaling}a has to be slightly modified: during backward evolution, the error gates occur before each evolution unitary gate instead of after it.}. We write the state $\rho^{\mathcal{N}_p}_{\psi}(t)$ as
\begin{eqnarray}
    \rho^{\mathcal{N}_p}_{\psi}(t) = q \ket{\psi_t}\bra{\psi_t} + (1-q)\tilde{\rho},
    \label{eq:dm_noisy}
\end{eqnarray}
where $\ket{\psi_t} = U_{\mathrm{Trotter}}^{t/\tau}(\tau) \ket{\psi}$ is the state after noiseless forward evolution and $q = (1 - p)^{ND / 2}$ is the probability that no error occurred during the forward evolution. The density matrix $\tilde{\rho}$ is the state conditioned on at least one error having occurred. The survival probability in noisy simulation then becomes
\begin{eqnarray}
    \begin{aligned}
        L_{A}^{\mathcal{N}_p}(t)
        = & q^2 |\braket{\psi_t | A | \psi_t }|^2 + (1-q)^2\tr \left[ (\tilde{\rho} A)^2 \right]\\
        &  + 2q(1-q) \braket{\psi_t | A \tilde{\rho} A | \psi_t}.
    \end{aligned}
    \label{eq:expansion_sp}
\end{eqnarray}
Defining $r = \sqrt{ \tr \left[\tilde{\rho}^2 \right] }$, we can use Cauchy-Schwarz inequality to obtain (see Appendix~\ref{app:proof})
\begin{eqnarray}
    \left | \frac{L_{A}^{\mathcal{N}_p}(t)}{q^2} - L_{A}(t) \right| \le (1-q)^2\left(\frac{r}{q}\right)^2 + 2(1-q)\frac{r}{q}.
    \label{eq:scaling_math}
\end{eqnarray}
Since $0 < q, r \le 1$, $L_{A}^{\mathcal{N}_p}(t) / q^2$ serves as a good approximation of $L_{A}(t)$ when $q \gg r$. This condition can be satisfied over a broad range of parameters because $r$ typically decays with the system size. In the most extreme case of global depolarizing noise, $\tilde{\rho}$ is a completely mixed state, for which $r^2 = 2^{-N}$. The condition $q \gg r$ then gives rise to
\begin{eqnarray}
    (1-p)^{ND} > \frac{C}{2^N} \Rightarrow  ND < \frac{N\log{2} + \log (1 / C)}{\log[1 / (1-p)]}
    \label{eq:limit_depth}
\end{eqnarray}
for some constant $C$. For $p=0.3\%$, this evaluates to $D < 230$ in the thermodynamic limit. For more general types of noise, we similarly expect the scaling with $q^2$ to hold up to some constant circuit depth in the thermodynamic limit. The noisy survival probability at this constant circuit depth will, however, decay exponentially when increasing the system size such that exponentially many measurements are required to resolve the signal. Nevertheless, we will show below that the number of measurements remains experimentally feasible in superconducting quantum devices for moderately sized systems with realistic error rates.

Two situations where Eq.~(\ref{eq:rescale}) fails directly follow from our argument. One is the case when $q$ approaches $r$, as already discussed. The other is when the initial state does not thermalize. For example, the product state $\ket{Z+} = \ket{0}^{\otimes N}$ is invariant under the (Floquet) XY Hamiltonian and thus will not get entangled. However, even in this case Eq.~(\ref{eq:rescale}) works well for many practical channels because two independent errors are unlikely to cancel each other.

\subsection{Numerical results}
\begin{figure}
    \centering
    \xincludegraphics[width=0.499\textwidth,trim={1.2cm 0.2cm 1.cm 0.cm},clip,label=(a)]{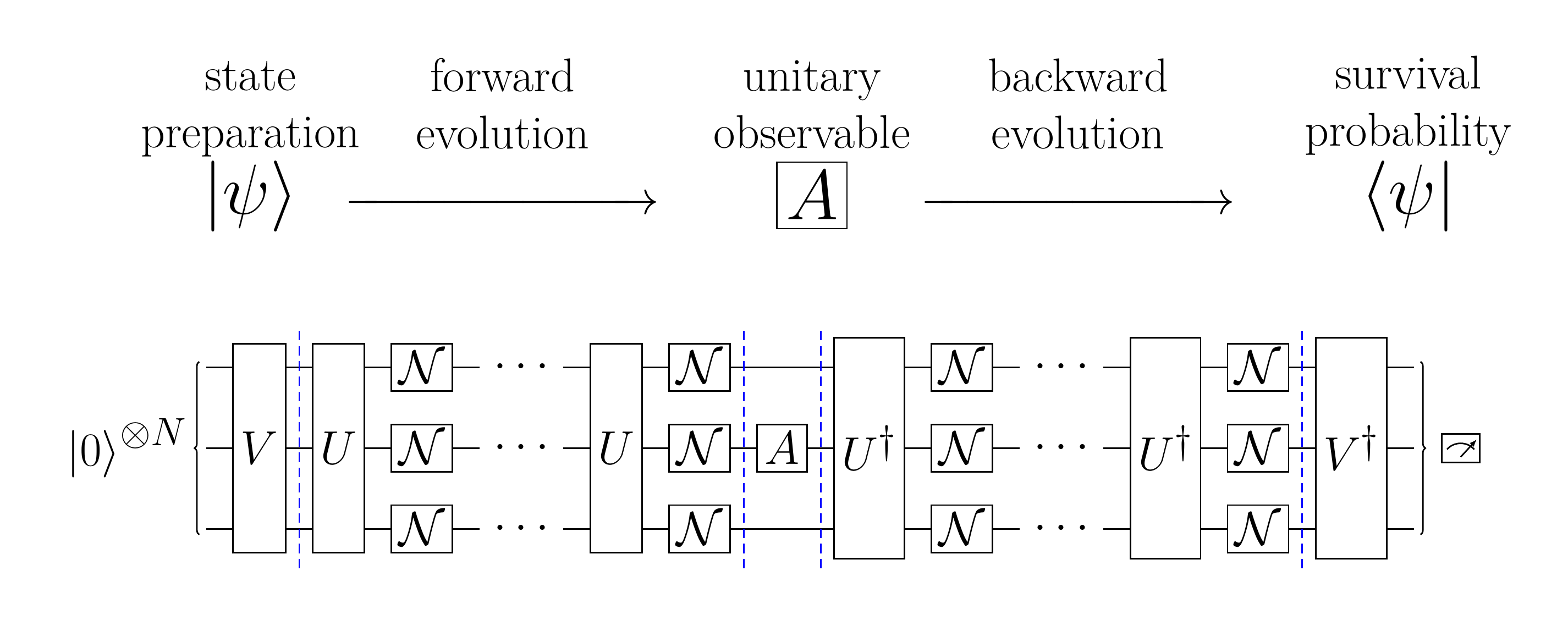}
    \\\ \\
    \xincludegraphics[width=0.25\textwidth, label=(b)]{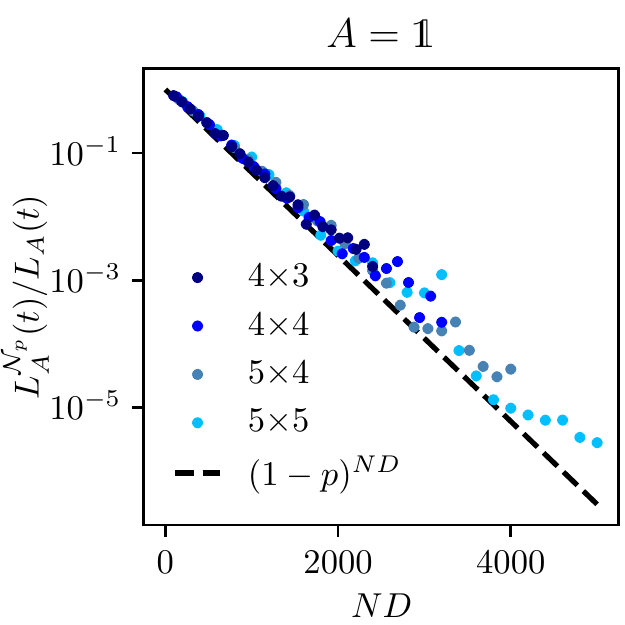}
    \xincludegraphics[width=0.2\textwidth, label=(c)]{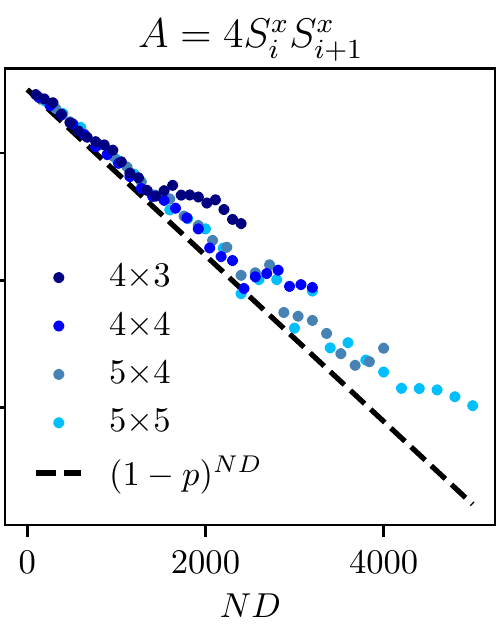}
    \caption{\textbf{(a)} Quantum circuit to map the expectation value of a (unitary) observable onto a survival probability. The initial state is prepared with $V$, $U= U_1, U_2, U_3$ or $ U_4$ is a single step in the Trotter decomposition, and $\mathcal{N}$ denotes a local noise channel.
    \textbf{(b)}, \textbf{(c)} Dependence of $L_{\id}^{\mathcal{N}_p}(t)$ and $L_{A}^{\mathcal{N}_p}(t)$ on the circuit depth $D$ and system size $N$ in the presence of depolarizing noise with error probability $p = 0.3\%$. The initial state is $\ket{\psi} = \ket{X+}$. The observable $A = 4S^x_{i}S_{i+1}^x$ is a correlator in the center of the lattice. The black dashed lines represent the scaling predicted by Eq.~(\ref{eq:scaling}).
    }
    \label{fig:scaling}
\end{figure}

We now numerically verify these considerations for the Floquet evolution of the XY model described in Sec.~\ref{sec:xy} in the presence of local depolarizing noise. For each qubit, the noise channel is given by
\begin{eqnarray}
    \mathcal{N}_{p}(\rho) = (1 - p)\rho + \sum_{\mu = 1}^{3} \frac{p}{3} \sigma^{\mu} \rho \sigma^{\mu}.
\end{eqnarray}
Other types of noise are discussed in the Appendix~\ref{sec:supp_plot}. In Fig.~\ref{fig:scaling}b and c, we respectively show $L_{\id}^{\mathcal{N}_p}(t)$ and  $L_{A}^{\mathcal{N}_p}(t)$ for the initial state $\ket{\psi} = \ket{X+}$ for different system sizes. The computations were performed using the Monte Carlo wavefunction method with the Cirq library~\cite{cirq_developers_2022_6599601}. Each data point in the figure corresponds to an average over 2000 quantum trajectories. This number of trajectories is sufficient to observe convergence of the mean value in the region of our interest. The results agree well with Eq.~(\ref{eq:scaling}). This also holds for different types of noise as we show in Appendix~\ref{sec:supp_plot}. We note that the data points start to deviate from the estimated black dashed lines at $ND$ approximately linear in $N$, in line with the expectation from Eq.~(\ref{eq:limit_depth}).

\begin{figure}
    \centering
    \xincludegraphics[width=0.235\textwidth, label=(a)]{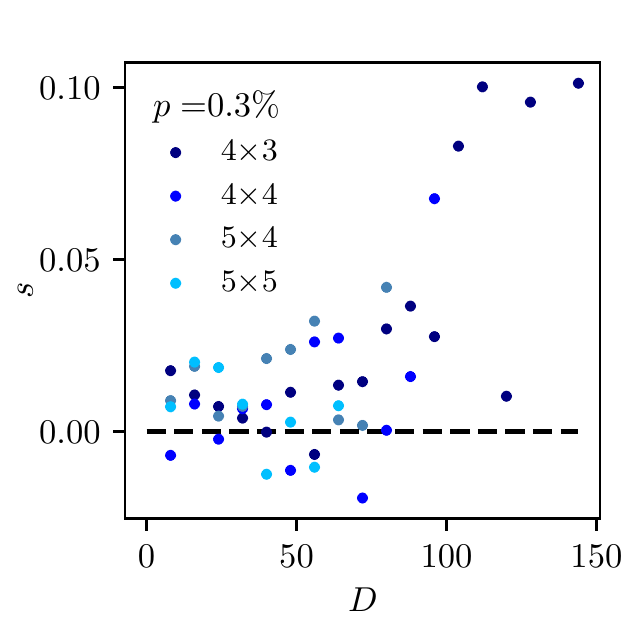}
    \xincludegraphics[width=0.235\textwidth, label=(b)]{figs/04b_TELoschmidt_alphaPi8.0_angle8_cutoff0.010_new_depolarizing.pdf}
    \caption{\textbf{(a)} The mitigation error $s_A^{\mathcal{N}_p}$ for the range of data in Fig.~\ref{fig:scaling} where $L_{\id}^{\mathcal{N}_p}(t) > 0.01$. We choose this cutoff due to the limited number of trajectories in the simulation, which limits the significant digits. \textbf{(b)} The root-mean-square of $s$, $\sqrt{\sum_{\mathrm{data}} s^2 / \sum_{\mathrm{data}}}$, evaluated over the window of circuit depth $[D - 16, D + 16]$ in \textbf{(a)}.}
    \label{fig:error_scaling}
\end{figure}

To quantify the error of the mitigation strategy, we define 
\begin{eqnarray}
    s_{A}^{\mathcal{N}_p}(t) = L_{A}^{\mathcal{N}_p}(t) \big/  L_{\id}^{\mathcal{N}_p}(t) - L_{A}(t).
\end{eqnarray}
Figure~\ref{fig:error_scaling}a shows the distribution of $s$ of the mitigated data from Fig.~\ref{fig:scaling}. The error remains small for depths up to $D \approx 100$. To compare different noise rates, we plot in Fig.~\ref{fig:error_scaling}b the square root of the moving average of $s^2$ for different values of $p$. Similar plots for types of noise other than depolarizing noise are presented in Appendix~\ref{sec:supp_plot}. For reference, the typical value of $L_{A}(t)$ in the simulation is around $0.3$, which indicates that for circuit depth $D=80$, the relative error is around 10\% for $p = 0.3\%$.

Although these results confirm the effectiveness of our error mitigation strategy, we also observe a systematic shift of $s$ towards positive values. This can be explained by the error terms in Eq.~(\ref{eq:expansion_sp}). Let us assume for simplicity that $\tilde{\rho} = \id / 2^N$, from which it follows that 
\begin{eqnarray}
    \begin{aligned}
    \frac{L_{A}^{\mathcal{N}_p}(t)}{L_{\id}^{\mathcal{N}_p}(t)} & = \frac{q^2 L_{A}(t) + (1 - q^2)/2^N}{q^2 + (1 - q^2)/2^N},
    \end{aligned}
\end{eqnarray}
where we used the fact that $A^2 = \id$ since $A$ is hermitian and unitary. Hence,
\begin{eqnarray}
    s_{A}^{\mathcal{N}_p}(t) = \left[1 - L_{A}(t)\right] \frac{(1 - q^2)}{q^2\cdot 2^N + (1 - q^2)} > 0.
    \label{eq:error_evaluation}
\end{eqnarray}
For certain error models, it may be possible to remove this systematic error by using a more complicated rescaling formula instead of \eqref{eq:rescale}. Nevertheless, the systematic error remains small as long as $q^2 \gg \tr(\tilde \rho^2)$.

\begin{figure}[t]
    \centering
    \includegraphics[width=0.38\textwidth]{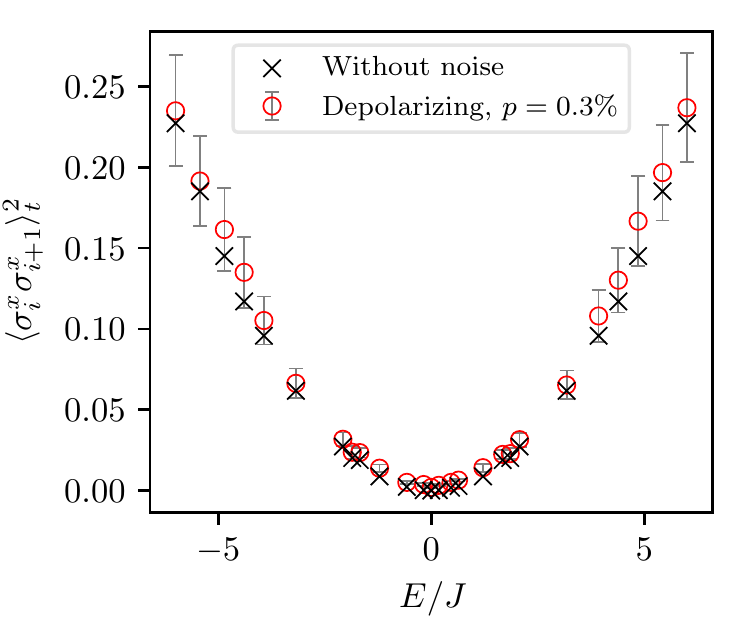}
    \caption{The time average of $L_{\sigma_i^x \sigma_{i+1}^x, \psi}(t)$ at $J t \approx 7.85$ with $\omega = 8J$, corresponding to 10 Trotter steps. The black crosses show the noiseless result. The red points were obtained by applying our error mitigation strategy to noisy simulations with a single-qubit depolarization rate $p = 0.3\%$. Error bars indicate the statistic errors due to fluctuations of different Monte Carlo trajectories, propagated from the standard deviations of $L_{A}^{\mathcal{N}_p}(t)$ and $L_{\id}^{\mathcal{N}_p}(t)$. The system size $N = 4 \times 4$.}
    \label{fig:full_exp} 
\end{figure}

We will now argue that our mitigation strategy enables the observation of prethermalization on current and near-term quantum devices. After Trotterization, the total required circuit depth $D$ to simulate time evolution of the two-dimensional XY model up to time $t_{\max}$ is
\begin{eqnarray}
	D = 4 \cdot 2 \cdot t_{\max} / \tau,
\end{eqnarray}
which, from left to right, represents the number of layers per Trotter step, back and forward evolution, and the number of Trotter steps. To see prethermalization of the Floquet XY model, Fig.~\ref{fig:prethermal} indicates that $t_{\max}$ should be at least $8 / J$ for $\omega = 8J$, which yields $D \approx 80$. The estimation is within the limit of the maximum circuit depth from Eq.~(\ref{eq:limit_depth}) and Fig.~\ref{fig:error_scaling} for $p =0.3\%$, showing that our proposal is suitable for current and near-term quantum devices.

We have now gathered all the ingredients for the full simulation of the PEVP on a noisy quantum device. We consider the two-dimensional XY model on a $4 \times 4$ square lattice in the presence of depolarizing noise with noise rate $p = 0.3 \%$. For the observable, we focus on the correlator $A = 4S_i^x S_{i+1}^x$ of a pair of neighboring sites at the center of the lattice.  In Fig.~\ref{fig:full_exp}, we plot the time averages of $\langle A(t) \rangle^2$ at driving frequency $\omega = 8J$ as a function of the initial state energy $E$ up to $t = 10 \tau$, corresponding to circuit depth $D=80$. The initial states were chosen from the same set as in Fig.~\ref{fig:prethermal}b. The black crosses represent the noise-free results, whereas for the red points the experiment was simulated including noise and error mitigation. The error bars show statistical errors due to fluctuations of different Monte Carlo trajectories, propagated from the standard deviations of $L_{A}^{\mathcal{N}_p}(t)$ and $L_{\id}^{\mathcal{N}_p}(t)$. Note that the sign of $\braket{\psi | A (t) | \psi }$ turns out to be constant during the Floquet time evolution in our range of simulations. In the long-time limit, the time average of the square is therefore equivalent to the square of the time average, given that they converge to a constant.

We find that the noise-free results lie within the error bars for all initial states and that the trend of the observable is well reproduced. This shows that our error mitigation procedure is viable to solve the PEVP. We note that the deviation between the noisy and noise-free results is biased since the red points are systematically above the black crosses, consistent with the expectation from Eq.~(\ref{eq:error_evaluation}).

\subsection{Implementation\label{sec:implementation}}

The results of the previous section show that our error mitigation strategy enables the solution of the PEVP for the XY model at a depolarizing noise rate of $p = 0.3 \%$. One more step remains to assess the experimental viability: an estimate of the number of required measurements.

In experiments, the survival probabilities are estimated from binary outcomes (success / failure). This gives rise to shot noise, which in turn sets a lower bound on the necessary number of samples. To achieve a statistical uncertainty of $\epsilon$, roughly $1/\epsilon^2$ samples are needed. For the error mitigation scheme to work, the shot noise must be smaller than the survival probability. As the noisy survival probability is suppressed by the factor $(1-p)^{ND}$, it follows that the number of needed measurements scales as $(1-p)^{-2ND}$. We note that this number of samples is typically orders of magnitude larger than the number needed to suppress the fluctuations in Monte Carlo trajectories due to noisy dynamics.

Since the sample complexity scales exponentially with the number of qubits, this is an important limitation to the system size that can realistically be reached. Nevertheless, classically hard regimes are accessible with realistic parameters. For instance, setting $N = 50$ while keeping $p = 0.3 \%$ and $D = 80$, we find that $(1-p)^{-2ND} \approx 3 \times 10^{10}$ samples are needed. This is inconveniently large as current superconducting quantum devices can collect millions of samples on the time scale of minutes. However, a modest improvement in the error rate to $p = 0.2\%$ reduces the number of samples to a much more realistic value of $9 \times 10^6$.

We have so far neglected the role of measurement errors, which occur with probability $p_{m} \approx 1 \% - 2\%$ for each single qubit measurement in current devices~\cite{Satzinger2021, Wei2022a}. Fortunately, these errors are automatically remedied by our error mitigation strategy. The measurement errors simply suppress the survival probability by another factor $(1-p_{m})^N$, which is independent of the circuit depth. For system sizes up to $N=50$, this increases the required number of measurements by at most an order of magnitude.

\section{Summary and outlook}
\label{sec:summary}

We have proposed the prethermal expectation value problem as a way to study thermal observables on noisy, intermediate-scale quantum devices. Our approach relies on the observation that relatively large Trotter steps, which do not permit a rigorous bound on the Trotter error, can give rise to prethermalization. We showed that in the prethermal regime, the equilibration of observables is similar to the expected dynamics under the original Hamiltonian. It may be possible to approximate evolution under the original Hamiltonian even better by cancelling higher-order terms of the Magnus expansion at the cost of more complex circuits. The range of energies at which the observables can be probed is set by the range of energies of the used intial states. We restricted ourselves to product states for this work, but the protocol can straightforwardly be extended to different initial states, which may increase the range of accessible energies.

We further demonstrated that the prethermal regime is experimentally accessible with noise rates of near-term devices using an error-mitigation scheme based on measuring and rescaling survival probabilities. This scheme is not limited to the PEVP but can be applied much more broadly in the context of quantum simulation. Our work provides all necessary ingredients to also study the approach to equilibrium and to extract, for instance, diffusion constants. Alternatively, one could consider the quantum dynamics of models which do not thermalize, such as quantum scars~\cite{Turner2018, Lin2019} or many-body localized systems~\cite{Pal2010, Abanin2018}.

Our work creates a new avenue to demonstrating useful quantum advantage on noisy devices. Although the XY model studied here can be efficiently simulated on classical computers with quantum Monte Carlo methods~\cite{Ding1992}, our approach can be readily adapted to more complex Hamiltonians. As a simple modification of the XY model, one might consider adding a site-dependent sign to the interaction strength $J$. This renders classical simulation of this model much harder since it causes a sign problem in quantum Monte Carlo methods~\cite{Loh1990, Takasu1986, Hatano1992}. The complexity of our proposed approach to quantum simulation however remains unaffected by this modification. Hence, quantum advantage may be within reach for studying the equilibrium properties of Hamiltonians with a sign problem.

\section*{Acknowledgements}
TEO and VS thank Yaroslav Herasymenko, Robin Kothari and Rolando Somma for useful discussions. We acknowledge the support from the German Federal Ministry of Education and Research (BMBF) through FermiQP (Grant No. 13N15890) and EQUAHUMO (Grant No. 13N16066) within the funding program quantum technologies - from basic research to market. This research is part of the Munich Quantum Valley (MQV), which is supported by the Bavarian state government with funds from the Hightech Agenda Bayern Plus. YY was funded by a grant from Google Quantum AI. DSW has received funding from the European Union’s Horizon 2020 research and innovation programme under the Marie Sk{\l}odowska-Curie Grant Agreement No. 101023276.
The work was partially supported by the Deutsche Forschungsgemeinschaft (DFG, German Research Foundation) under Germany's Excellence Strategy -- EXC-2111 -- 390814868.

\bibliography{main}

\clearpage
\appendix
\begin{appendices}
    \section{Definition of problems}
\label{sec:def}

\subsection{Setup}
In this section we consider 
\begin{itemize}
    \item a local Hamiltonian $H$ as considered in \ref{sec:equilibration}, with spectral decomposition
        \begin{eqnarray}
            H = \sum_k E_k \ket{k}\bra{k},
        \end{eqnarray}
    \item the Trotterized time-evolution unitary $U_{\mathrm {Trotter}}(\tau)$ (see Eq.\ \ref{def:Utrot}) with time step $\tau$,
    \item an observable $A$ with operator norm $\Vert A \Vert$,
    \item and an initial state $\ket{\psi}$.
\end{itemize}
When also given a Trotter step $\tau$, any time appearing in text will be stroboscopic, i.e., an integer multiple of $\tau$.

\subsection{Definition of thermal ensembles} Here we provide definitions of the microcanonical and diagonal ensembles in Fig.~\ref{fig:prethermal}.
\begin{definition}[The microcanonical ensemble]
    \label{def:micro}
    Given an energy $E$ and energy interval $\delta$, the value of an observable $A$ in the corresponding microcanonical ensemble is defined as
    \begin{equation}
        \langle A \rangle_{\mathrm{micro}, E} = \sum_{k \in I_{E,\delta}} \braket{k | A | k}  / |I_{E, \delta}|,
    \end{equation}
    where $I_{E,\delta} = \left\{ k | |E_k - E| < \delta / 2 \right\}$.
\end{definition}
Alternatively, for the convenience of computation, the energy cutoff may be replaced by a Gaussian filter:
\begin{definition}[The broadened microcanonical ensemble]
    With the same setup as Definition~\ref{def:micro}, the broadened microcanonical ensemble is defined as
    \begin{equation}
        \langle A \rangle_{\mathrm{micro}^{\prime}, E} = \sum_k \braket{k | A | k} e^{-\frac{(E - E_k)^2}{2\delta ^2}} \big/ \sum_k e^{-\frac{(E - E_k)^2}{2\delta ^2}}.
    \end{equation}
\end{definition}
 The two definitions are equivalent in the thermodynamic limit under the eigenstate thermalization hypothesis~\cite{Lu2020, Yang2022}. In Fig.~\ref{fig:prethermal} we take the latter definition, which can be efficiently computed in 1D systems with classical computers using filtering algorithms for $\delta$ being a constant~\cite{Yang2022}.

\begin{definition}[The diagonal ensemble]
    Given a state $\ket{\psi}$, the value of an observable $A$ in the the diagonal ensemble is defined as
    \begin{equation}
        A_{d,\psi} = \sum_k |\langle \psi | k \rangle|^2 \langle k | A | k \rangle.
    \end{equation}
\end{definition}
The diagonal ensemble values are equivalent to the long time average of the initial state $\ket{\psi}$ and observable $A$ for non-degenerate Hamiltonians. It can be approximated again by filtering out the off-diagonal elements of an initial density matrix~\cite{Cakan2021}. The entanglement entropy of the diagonal ensemble in operator space however obeys a volume law scaling, which limits the system size reachable in classical simulations.

\subsection{Definition of PEVP} To define the PEVP, we first need give a precise definition of a prethermal plateau. 
There is not a single accepted definition for a prethermal plateau in the literature. 
Here we formulate the practical definition we use. First we define what we consider to be a plateau.

\begin{definition} [The plateau]
    \label{def:plateau}
    Given a tolerance $\epsilon \ll 1$, a plateau is a time interval $[t_1, t_2)$ with $t_1 < t_2 \le \infty$ such that
    \begin{enumerate}
        \item $\max\limits_{t_1 \leq t < t_2} \braket{A}_t - \min\limits_{t_1 \leq t < t_2} \braket{A}_t \leq \epsilon \Vert A\Vert$,
        where $\langle A \rangle_t$ is defined in Eq.~(\ref{eq:AFloq}).
        \item there exists no overlapping interval $[t_1^{\prime}, t_2^{\prime})$ also satisfying 1 for which $t_2^{\prime} / t_1^{\prime} > t_2 / t_1$.
    \end{enumerate}
\end{definition}
The second criterion ensures the plateau we find is locally the longest. Here we take the ratio $t_2/t_1$ as the measure of the length of the plateau to be more consistent with the ideas of prethermalization.  A plateau can be identified as a prethermal plateau, if 
\begin{itemize}
\item it is not connected to the final Floquet thermalization plateau at infinite time and temperature~\cite{Mori2018},
\item the ratio $t_2 / t_1$ grows exponentially with $1 / \tau$ and
\item in the small $\tau$ limit, $t_1$ converges to a positive number.
\end{itemize}

It is in general hard to identify a prethermal plateau, due to the difficulty of reaching the exponentially growing $t_2$ in simulations. Nevertheless, assuming its existence, it is relatively easy to find the plateau and compute the plateau value. Now let us restate Problem~\ref{prob:PEVP} in the main text:
\begin{definition}[The prethermalized expectation value problem]
    Given a unitary $U_{\mathrm{Trotter}}(\tau)$, a state $\ket{\psi}$, and a local observable $A$, assume that a prethermal plateau exists between times $t_1$ to $t_2$, such that $\max_{t \in [t_1, t_2)} \langle A \rangle_t - \min_{t \in [t_1, t_2)} \langle A \rangle_t \leq \epsilon \Vert A\Vert$. Find the value of $\braket{A}_t$ to within additive error $2 \epsilon \Vert A\Vert$ for any $t \in [t_1, t_2)$ .
\end{definition}

\section{The Magnus expansion}
\label{sec:magnus}

The Magnus expansion serves as a series expansion for the effective Hamiltonian of a Floquet driving $H(t)$ with period $\tau$:
\begin{eqnarray}
	\begin{aligned}
		{U}_{F}(\tau) = & \mathcal{T}\left( e^{-i \int_{0}^{\tau}  {H}(t) \dInt t} \right) = e^{-i \tau \sum_{k=1}^{\infty} \Omega_k}                       \\
		\Omega_0 =     & \frac{1}{\tau} \int_{0}^{\tau} \dInt t_1 {H}(t_1)                                                                                                 \\
		\Omega_1 =                   & \frac{1}{2i \tau} \int_{0}^{\tau} \dInt t_1 \int_{0}^{t_1} \dInt t_2 \left[{H}(t_1), {H}(t_2)\right]                                 \\
		\Omega_2 =                   & - \frac{1}{6\tau} \int_{0}^{\tau} \dInt t_1 \int_{0}^{t_1} \dInt t_2 \int_{0}^{t_2} \dInt t_3 \\
        ([{H} & (t_1), [{H}(t_2), {H}(t_3)]]  + [{H}(t_3), [{H}(t_2), {H}(t_1)]])                                                                                        \\
		                                   & \vdots
	\end{aligned}
\end{eqnarray}

In general, the Magnus expansion is not convergent~\cite{Blanes2009, Bukov2015} and thus higher order contributions are not negligible for finite driving frequencies. Nevertheless, its finite truncation is still expected to approximate the quasi-stationary prethermal plateau~\cite{kuwahara2016}. To be more precise, let ${H}_{\mathrm{Magnus}}^{(n)} = \sum_{j = 0}^{n} \Omega_j$ denote the $n$-th order truncated effective Hamiltonian, then there exists $n_0 = \varO(\omega / kJ)$ such that 
\begin{eqnarray}
    \Vert U_F(\tau)^m - e^{-i{H}_{\mathrm{Magnus}}^{(n_0)} m\tau} \Vert \lesssim N m \tau 2^{-n_0}.
    \label{eq:floq_magnus1}
\end{eqnarray}

The general estimation Eq.~(\ref{eq:floq_magnus1}) for the unitary evolution operators has a linear dependence on system size, which does not imply prethermalization for $N \gtrsim \exp(\varO(\omega / kJ))$. When considering local observables acting on a subsystem $L$ and short-range interacting Hamiltonians, however, the bound can be tightened for the reduced density matrix $\rho_L$:
\begin{eqnarray}
    \begin{aligned}
        \Vert (\rho_L)_F(m\tau) -  (\rho_L)_{\mathrm{Magnus}}^{(n_0)}(m\tau) \Vert_1
        \lesssim |L|m\tau e^{- \varO(\omega)}
    \end{aligned}
\end{eqnarray}
for the same $n_0$, where the system size dependence is erased~\cite{kuwahara2016}. 

\begin{figure}
    \centering
    \xincludegraphics[width=0.38\textwidth,trim={0.6cm 0.5cm 0.6cm 1cm},clip, label=(a)]{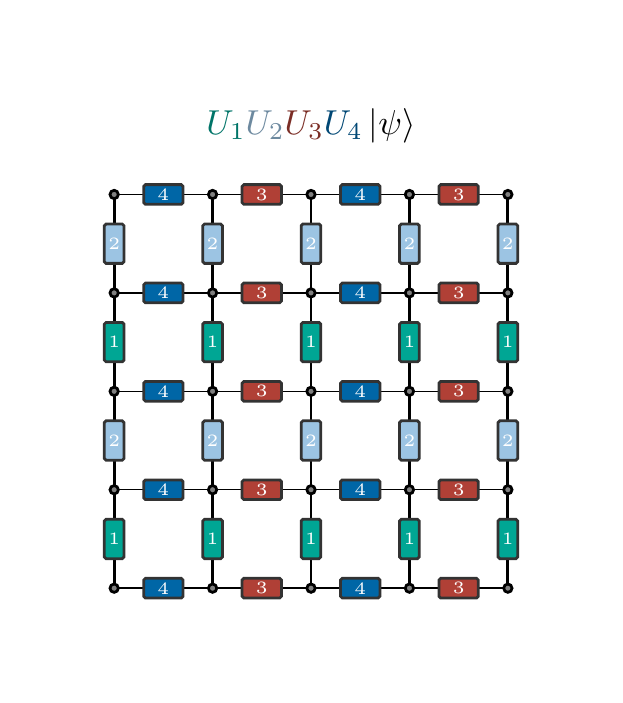}
    
    \xincludegraphics[width=0.235\textwidth, label=(b)]{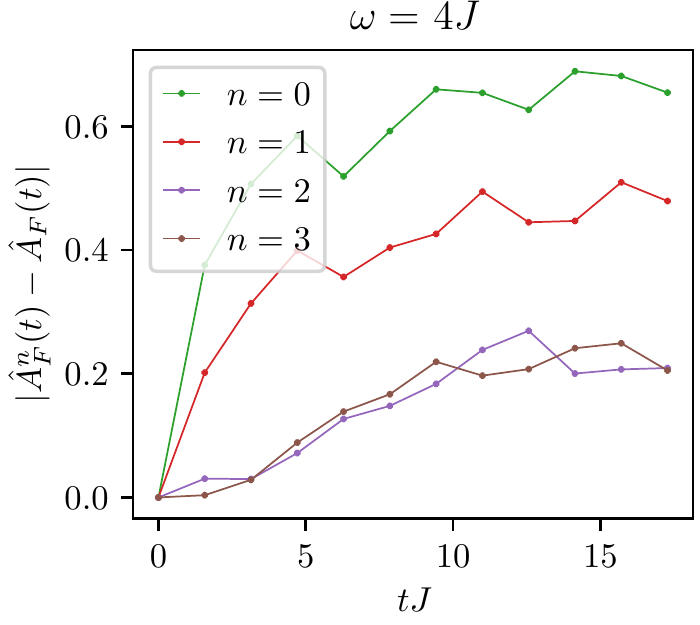}
    \xincludegraphics[width=0.235\textwidth, label=(c)]{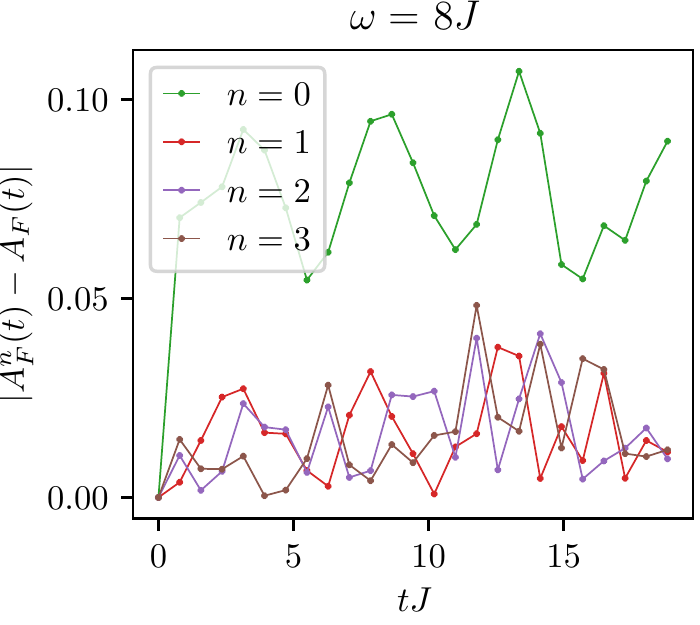}
    \caption{\textbf{(a):} Trotterization of 2D XY Hamiltonian. The circles represent qubits and the rectangles on the bounds represent two-qubit evolution gates on neighbouring qubits. 
    \textbf{(b-c):} Simulation of evolving Floquet XY model with the Magnus expansion truncated to $n$-th order. Their differences from Floquet evolution are plotted. The initial state is $\ket{X+}$ and ${A} = m_x^2 + m_y^2$. \textbf{(b):} $\omega = 4J$, \textbf{(c):} $\omega = 8J$. Note the difference in scale on the y-axis. The system size is $N = 4 \times 3$.}
    \label{fig:magnus}
\end{figure}

For the proof of this relation to hold rigorously, the required driving frequency is $\omega \ge 16\pi kJ \approx 100 J$ for nearest neighbour interacting Hamiltonians, while in our numerical simulation in Fig.~\ref{fig:prethermal}, prethermalization has occurred for $\omega \sim 8 J $. For all of our numerical simulations of the XY-model, we use the Trotterization shown in Fig.~\ref{fig:magnus}a. In Fig.~\ref{fig:magnus}b-c the differences between Floquet evolution and its Magnus expansions up to the third order are plotted. Note that the zeroth order Magnus expansion is just the original non-Floquet Hamiltonian. For $\omega = 8J$, it turns out that the $n = 1$ case already gives a good approximation of the Floquet Hamiltonian.

\section{Difficulty of error mitigation in time evolution}
\label{sec:diff_TE}
The difficulty of error mitigation of observables by measuring them directly can be explained in the following two ways. 

First, if we take the formalism as in Eq.~(\ref{eq:dm_noisy}), the aim will be to obtain $A_{\psi}(t) = \braket{\psi_t | A | \psi_t}$ from
\begin{eqnarray}
    A_{\psi}^{\mathcal{N}_p}(t) = \tr \left( A \rho_{\psi}^{\mathcal{N}_p} \right) = qA_{\psi}(t) + (1-q) \tr\left( 
A \tilde{\rho} \right).
\label{eq:direct_te}
\end{eqnarray}
Although the second term vanishes for global depolarizing channel and traceless $A$, one can not use the same trick as Eq.~(\ref{eq:rescale}) to directly estimate $q$, since setting $A = \id$ would not give any meaningful output. Of course, it is in principle still possible to measure the survival probability with backward evolution that approximates $q^2$ and take its square root. In the latter circuit, however, any coherent noise will partially cancel in forward and backward evolutions, which gives a different value of $q$ from the one we need in Eq.~(\ref{eq:direct_te}).

Alternatively, we can think about the problem using a random walk picture, where an initial state will be quickly heated during time evolution on noisy digial simulators, because of the strong energy dependence of the density of states (DOS). Let us consider the quantum trajectory simulation process of a noisy circuit. Assume the absolute average energy change per error to be a constant $g > 0$ and denote the expectation value of the energy of the simulated state after $n$ errors by $E_n$. The probability of increasing or decreasing energy after each gate of noise will be
\begin{eqnarray}
  \frac{\mathbb{P}(E_{n+1} = E_n + g)}{\mathbb{P}(E_{n+1} = E_n -g)} = \frac{\mathrm{DOS}(E_n+g)}{\mathrm{DOS}(E_n-g)}.
\end{eqnarray}
For short-range interacting and locally bounded Hamiltonians, the DOS converges weakly to a Gaussian in the thermodynamic limit~\cite{Hartmann2005}:
\begin{eqnarray}
    \mathrm{DOS}(E) \propto \exp\left( - E^2 / 2N\sigma^2 \right),
\end{eqnarray}
where $\sigma$ is a constant depending on local energy scale. Inserting ${\mathbb{P}(E_{n+1} = E_n + g)} + {\mathbb{P}(E_{n+1} = E_n -g)} = 1$, it can be concluded that
\begin{eqnarray}
    \begin{aligned}
      \Delta E =  & E_{n+1} - E_n\\
      = & g \left[\mathbb{P}(E_{n+1} = E_n + g) - \mathbb{P}(E_{n+1} = E_n -g) \right]                  \\
      = & -g \tanh \left(\frac{gE_n}{N\sigma^2}\right)
    \end{aligned}
\end{eqnarray}
The circuit depth $D$ required for a single noise to occur is $\Delta D = 1 / p N$, where $p$ is the noise rate. Therefore
\begin{eqnarray}
    \frac{\Delta E}{\Delta D} = - p g N \tanh \frac{gE}{N\sigma^2},
\end{eqnarray}
whose solution in the continuous limit is
\begin{eqnarray}
    \sinh \left( \frac{g}{N\sigma^2}E \right) = \sinh \left( \frac{g}{N\sigma^2}E_0 \right) e^{-pg^2D / \sigma^2}.
\end{eqnarray}
It gives rise to an exponential decay in energy with regard to the circuit depth. In other words, the initial state will be heated to infinite temperature, and this process is much faster than the heating caused by Floquet driving in the prethermal regime. Post-selection error mitigation strategies for direct time evolution would then imply that it is possible to extract low temperature properties from higher temperatures. There is no reason to assume that this would be the case, especially in the case when phase transitions exist.

\begin{figure}[b]
    \centering
    (a) Phase damping noise\\
    \xincludegraphics[width=0.25\textwidth, label=(a1)]{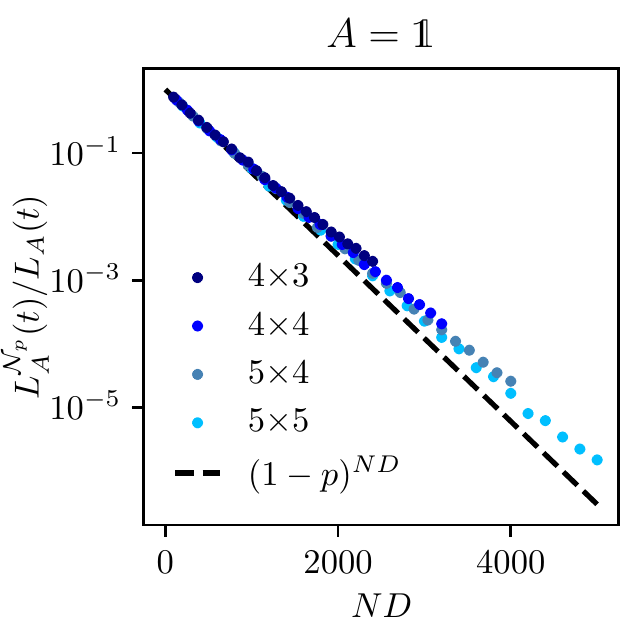}
    \xincludegraphics[width=0.2\textwidth, label=(a2)]{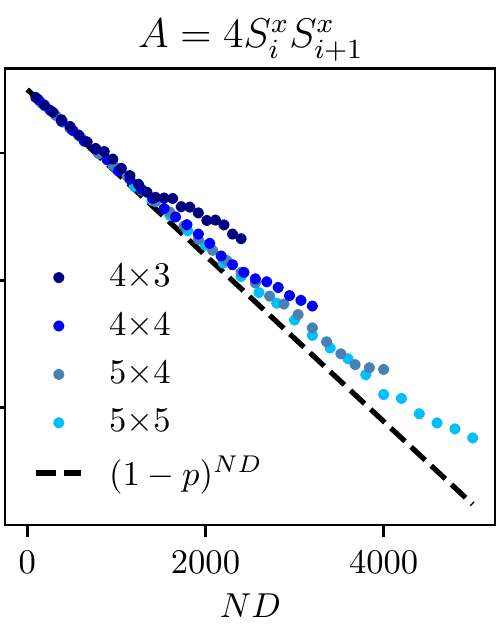}
    \xincludegraphics[width=0.235\textwidth, label=(a3)]{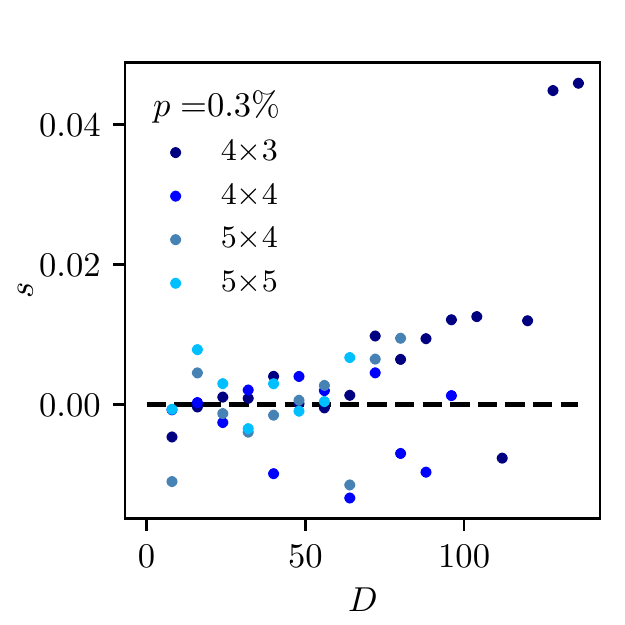}
    \xincludegraphics[width=0.235\textwidth, label=(a4)]{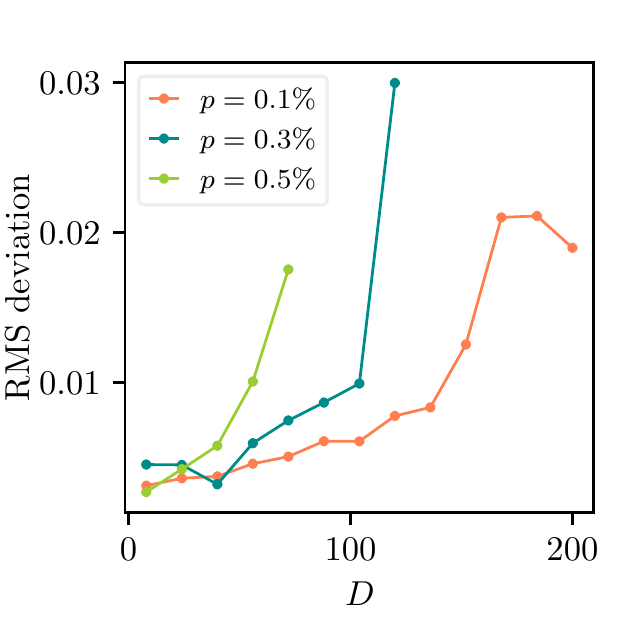}
    (b) Amplitude damping noise\\
    \xincludegraphics[width=0.25\textwidth, label=(b1)]{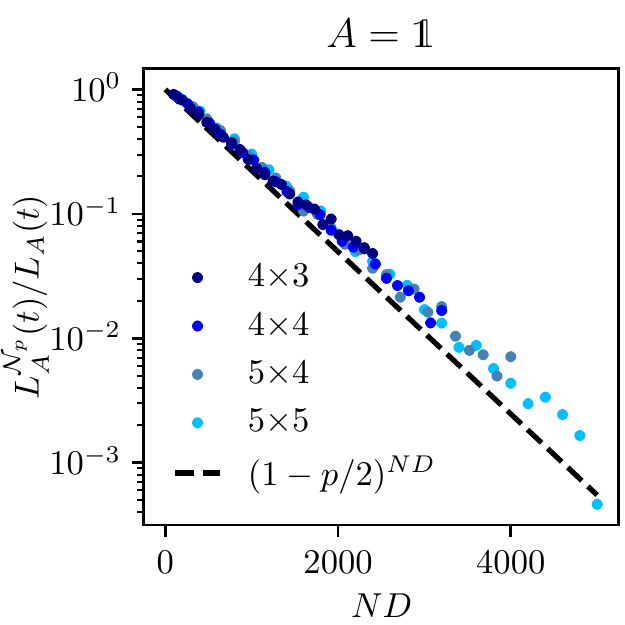}
    \xincludegraphics[width=0.2\textwidth, label=(b2)]{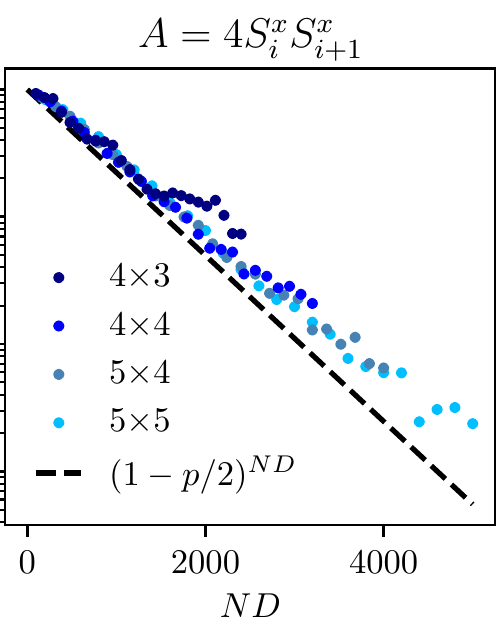}
    \xincludegraphics[width=0.235\textwidth, label=(b3)]{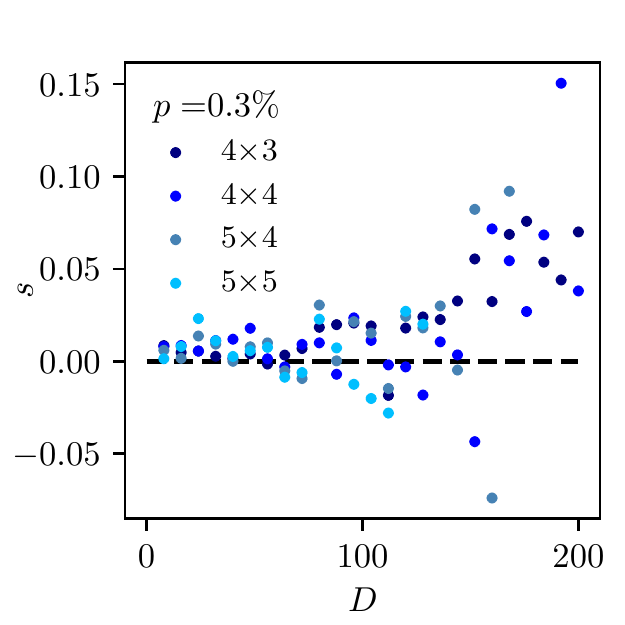}
    \xincludegraphics[width=0.235\textwidth, label=(b4)]{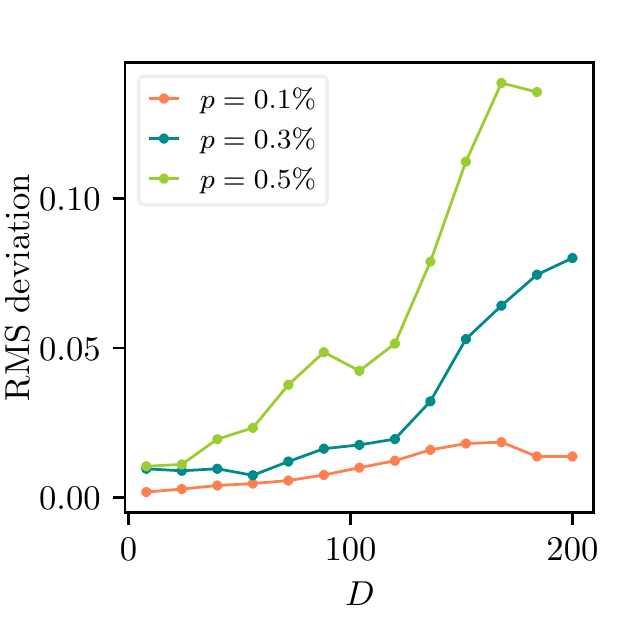}
    \caption{Simulation results for phase damping (a1-a4) and amplitude damping (b1-b4) noise and $p=0.3\%$. The plotted quantities are the same as shown in Fig.~\ref{fig:scaling} and Fig.~\ref{fig:error_scaling}.}
    \label{fig:supp_plot}
\end{figure}

\section{Phase and amplitude damping noises}
\label{sec:supp_plot}

In the main text, we focused on depolarizing noise. In this appendix, we show that the effects of phase damping and amplitude damping noise are qualitatively similar. The relevant noise channels are given by~\cite{Nielsen2012}:
\begin{itemize}
    \item the phase damping channel
    \begin{eqnarray}
        \mathcal{N}_{p}^{P}(\rho) = (1 - p) \rho + p \sigma_i^z \rho \sigma_i^z,
    \end{eqnarray}
    \item and the amplitude damping channel
    \begin{eqnarray}
        \mathcal{N}_{p}^{A}(\rho) = M_0 \rho M_0^{\dagger} + M_1 \rho M_1^{\dagger},
    \end{eqnarray}
    where $M_0 = \begin{pmatrix}
        1 & 0 \\ 0 & \sqrt{1 - p}
    \end{pmatrix}$ and $M_1 = \begin{pmatrix}
        0 & \sqrt{p} \\ 0 & 0
    \end{pmatrix}$.
\end{itemize}

In Fig.~\ref{fig:supp_plot} we plot the simulation results for these two types of noises in the same fashion as in Fig.~\ref{fig:scaling} and Fig.~\ref{fig:error_scaling}. From top to bottom, they are the scaling of survival probability without (left) and with (right) applying the observable, the error $s$ of the mitigation strategy and the moving quadratic average of $s$. The scalings are also fit well with Eq.(\ref{eq:scaling}), while the error after rescaling is much smaller for phase damping error than for the other two. Note that for amplitude damping noise, the effective survival probability is $q^2 = (1 - p / 2)^{ND}$. This is likely due to the balanced distribution of our initial states in the $z$ direction, which reduces the probability of seeing a single state jumping to $p/2$.

\begin{figure}
    \centering
    \xincludegraphics[width=0.25\textwidth, label=(a)]{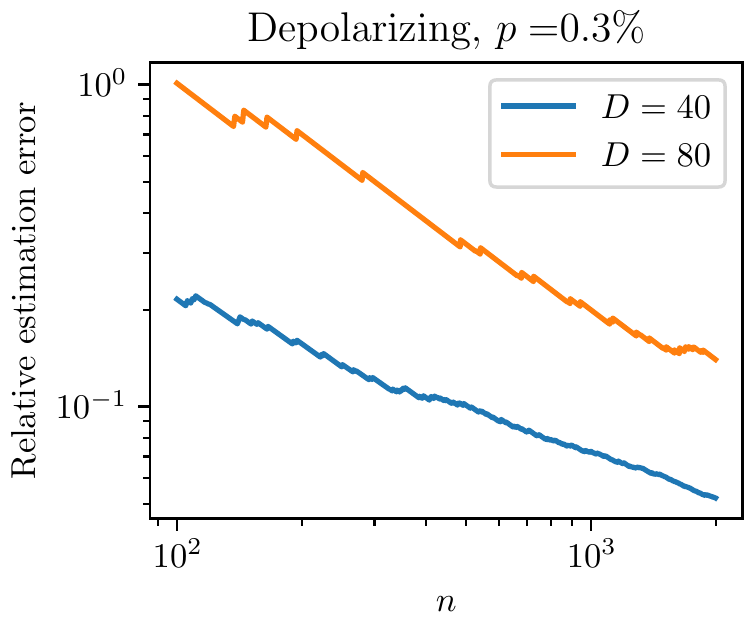}
    \\
    \xincludegraphics[width=0.235\textwidth, label=(b)]{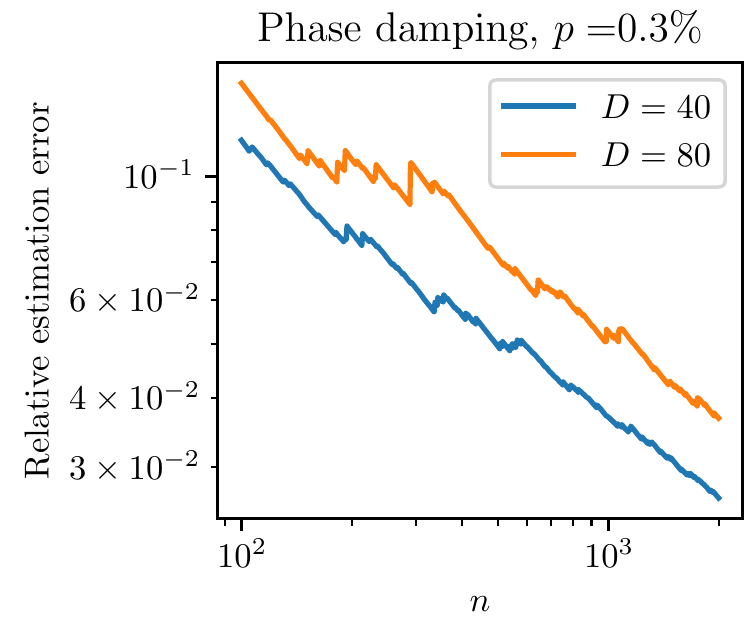}
    \xincludegraphics[width=0.235\textwidth, label=(c)]{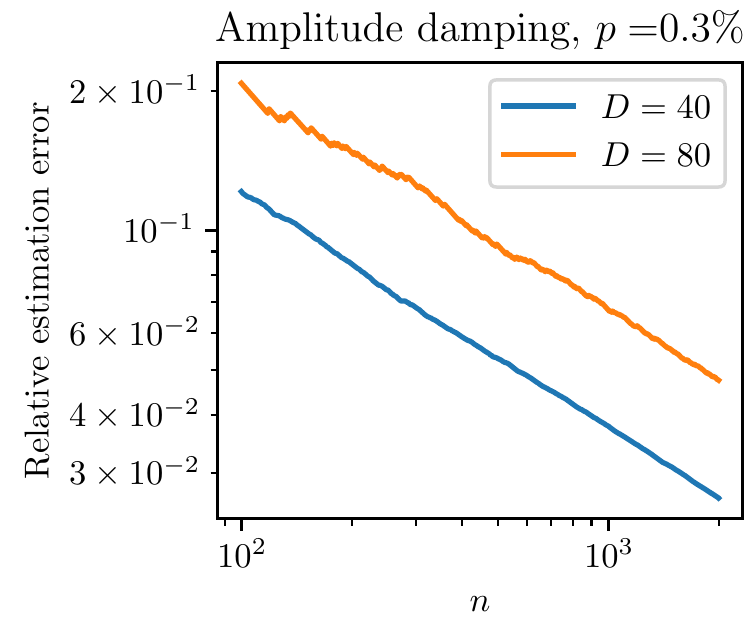}
    \caption{The relative estimation error of $L_{\id}^{\mathcal{N}_p}(t)$ as a function of the number $n$ of Monte Carlo samples for circuit depth $D=40$ and $80$. The estimation error is defined as the standard deviation of the ensemble of expectation values from the trajectories divided by $\sqrt{n}$. The figures show the relative estimation error, i.e.~the ratio of the estimation error to the estimated value (mean) of $L_{\id}^{\mathcal{N}_p}(t)$. The system size is $4\times 4$ and the initial state is $\ket{X+}$.}
    \label{fig:convergence}
\end{figure}

In Fig.~\ref{fig:convergence}, we show the convergence of the Monte Carlo simulations of $L_{\id}^{\mathcal{N}_p}(t)$ . We observe that phase and amplitude damping noises require a much smaller number of trajectories than depolarizing noise to reach the same estimation error. For circuit depth $D=80$ and $n=2000$ trajectories, which are the parameters used in Fig.~\ref{fig:full_exp}, the error can be read off from Fig.~\ref{fig:convergence}a) to be about $15\%$.

\section{Phase and amplitude damping noises}
\label{sec:supp_plot}

\section{Proof of Eq.~(\ref{eq:scaling_math})\label{app:proof}}
The trace of the product of two matrices $\tr \left( A^{\dagger} B \right)$ can be viewed as an inner product, and thus the Cauchy-Schwarz inequality applies:
\begin{eqnarray}
    \left|\tr \left( A^{\dagger} B \right) \right| \le \sqrt{\tr \left( A^{\dagger} A\right) \cdot \tr \left( B^{\dagger} B \right)}.
\end{eqnarray}
Since ${A}$ is hermitian and  unitary, ${A}^2 = \id$ and the first perturbation term in Eq.~(\ref{eq:expansion_sp}) can be bounded by
\begin{eqnarray}
    \begin{aligned}
        \left|\tr \left[ (\tilde{\rho} {A})^2 \right]\right|          
        \le  \sqrt{\tr \left[({A} \tilde{\rho} {A})^2\right] \cdot \tr \left[\tilde{\rho} ^2\right]}
        =   r^2.
    \end{aligned}    
\end{eqnarray}
Similarly, for the other term,
\begin{eqnarray}
    \begin{aligned}
        & |\braket{\psi_t | {A} \tilde{\rho} {A} | \psi_t}|  \\
        = & \left|\tr \left( {A} \tilde{\rho} {A} \ket{\psi_t}\bra{\psi_t} \right) \right| \le \sqrt{\tr \left[({A} \tilde{\rho} {A})^2\right]} = r.
    \end{aligned}
\end{eqnarray}
Combining these inequalities, we get Eq.~(\ref{eq:scaling_math}).
\end{appendices}

\end{document}